\documentclass[twocolumn,trackchanges]{aastex701}

\shorttitle{Black Hole Occupation Fraction}
\shortauthors{Contini et al.}

\begin{document}

\title{Black Hole Occupation Fraction: Dependence on Black Hole Mass Threshold, Environment, Resolution and Redshift.}

\author{Emanuele Contini}
\altaffiliation{Yonsei University}
\affiliation{Department of Astronomy and Yonsei University Observatory, Yonsei University, 50 Yonsei-ro, Seodaemun-gu, Seoul 03722, Republic of Korea}
\email[show]{emanuele.contini82@gmail.com}

\author{J.K. Jang}
\altaffiliation{Yonsei University}
\affiliation{Department of Astronomy and Yonsei University Observatory, Yonsei University, 50 Yonsei-ro, Seodaemun-gu, Seoul 03722, Republic of Korea}
\email{starbrown816@gmail.com}

\author{Jinsu Rhee}
\altaffiliation{Institut d’Astrophysique de Paris}
\affiliation{Institut d’Astrophysique de Paris, Sorbonne Université, CNRS, UMR 7095, 98 bis bd Arago, 75014 Paris, France}
\email{emanuele.contini82@google.com}

\author{Changjo Seo}
\altaffiliation{Yonsei University}
\affiliation{Department of Astronomy and Yonsei University Observatory, Yonsei University, 50 Yonsei-ro, Seodaemun-gu, Seoul 03722, Republic of Korea}
\email{emanuele.contini82@gmail.com}

\author{Sukyoung K. Yi}
\altaffiliation{Yonsei University}
\affiliation{Department of Astronomy and Yonsei University Observatory, Yonsei University, 50 Yonsei-ro, Seodaemun-gu, Seoul 03722, Republic of Korea}
\email[show]{yi@yonsei.ac.kr}

\begin{abstract}
We take advantage of the state-of-the-art semi-analytic model \texttt{FEGA25} \citep{contini2025}, run on merger trees extracted from three dark matter-only cosmological simulations, to study the relation between the black hole (BH) occupation fraction, $f_{\rm BH,occ}$, and galaxy stellar mass as a function of BH mass threshold, galaxy type, simulated volume, numerical resolution, sampled galaxy population, and redshift. \texttt{FEGA25} includes an improved treatment of active galactic nucleus feedback and does not impose a pre-existing BH seed population: BHs grow naturally through quasar and radio modes. Starting from the prerequisite that \texttt{FEGA25} reproduces the observed BH mass function from at least $z=2$ to the present day, our analysis leads to several results. We find that $f_{\rm BH,occ}$ increases with stellar mass, but that its normalization and shape depend strongly on the adopted BH mass threshold and on the relative contribution of central and satellite galaxies. The relative behavior of central and satellite galaxies depends on the simulation box and BH mass threshold, while the global relation should be interpreted as a population-weighted quantity. We also find significant box-to-box variations, reflecting the combined impact of numerical resolution, simulated volume, and sampled galaxy population. The redshift evolution is not universal: YS50 and the \texttt{NewCluster} zoom-in simulation show a trend qualitatively similar to that reported by \citet{tremmel2024}, whereas larger-volume boxes show the opposite behavior. Finally, comparison with other studies shows that the inferred occupation fraction is highly sensitive to BH mass threshold, simulated
volume, numerical resolution, and sampled galaxy population.
\end{abstract}



\section{Introduction}
\label{sec:intro}

The growth of black holes (BHs) at the centers of galaxies is still a matter of debate. These BHs can be massive, $M_{\rm BH}>10^5\,M_{\odot}$, and both their formation and evolution remain controversial, mainly because of the lack of sufficiently strong observational constraints. Recent observations have shown that massive BHs were already in place at very early cosmic times. In faint JWST-selected galaxies, accreting BHs with masses of order $\sim10^{6}-10^{7}\,M_{\odot}$ have been inferred at $z\gtrsim8-10$ \citep[e.g.,][]{larson2023,maiolino2024,natarajan2024}. At the same time, luminous quasars at $z\sim7.5$ are known to host BHs with masses $\gtrsim10^{9}\,M_{\odot}$ \citep[e.g.,][]{banados2018,wang2021}. These discoveries indicate that BH formation and growth must have proceeded very efficiently during the first few hundred million years of cosmic history.

These recent observational discoveries pose challenging problems for theoretical models. In order to form such massive BHs at early epochs, models generally require efficient initial seeding and/or rapid subsequent growth (perhaps in super-Eddington accretion). There are mainly two possible formation channels for BH seeds: \emph{light} seeds with $M_{\rm BH}\sim10^2\,M_{\odot}$, originating from the first generation of Population III stars (e.g., \citealt{madau-rees2001,taylor-kobayashi2014,smith2018,wise2019}), and \emph{heavy} seeds with $M_{\rm BH}\sim10^{4-6}\,M_{\odot}$, possibly produced by the direct collapse of massive gas clouds (e.g., \citealt{volonteri2008,begelman2010,alexander-natarajan2014,ferrara2014,wise2019,begelman-silk2023}). Both light and heavy seeds may contribute to the observed BH population, possibly with different relative importance across cosmic time. It is also possible that neither channel alone provides a complete description of the true BH formation pathway.

What is clear is that massive BHs are observed at high redshift, and a purely light-seed scenario may be insufficient unless very efficient growth is invoked. This does not imply that BHs cannot form from Population III remnants and subsequently grow through accretion and mergers, but it does highlight the need for additional constraints. One way to characterize the relative importance of different formation scenarios is through the BH occupation fraction, defined as the fraction of galaxies, at fixed stellar mass, that host a BH above a given mass threshold, typically above $M_{\rm BH} \sim 10^5 M_{\odot}$ (e.g., \citealt{miller2015,bellovary2019,haidar2022,tremmel2024,bhowmick2025}). This quantity is particularly informative in the dwarf-galaxy regime, $M_*<10^{9.5-10}\,M_{\odot}$. In the local Universe, galaxies above this mass range are known to have occupation fractions close to $100\%$ (e.g., \citealt{miller2015,tremmel2024}). By contrast, the BH occupation fraction in dwarf galaxies remains highly uncertain. Since these systems do not grow substantially through either accretion or mergers, and neither do their BHs, they may preserve memory of the initial conditions of BH formation \citep{tremmel2024}. Studying the BH occupation fraction in this mass range can therefore shed light on BH seeding mechanisms \citep{volonteri-natarajan2009,greene2012,tremmel2024}.

Many studies have focused on the importance of BH seeding (\citealt{bhowmick2025} and references therein) and on the shape and normalization of the BH occupation fraction as a function of stellar mass, both theoretically and observationally (e.g., \citealt{tremmel2024,bhowmick2025,zou2025}), with particular attention to the dwarf-galaxy regime. A key message from \citet{haidar2022}, who compared a wide range of predictions from numerical simulations, is that theoretical estimates of the BH occupation fraction can differ substantially because of differences in seeding prescriptions, accretion models, feedback implementations, and numerical resolution. Consistently with this picture, \citet{bellovary2019} found that the occupation fraction increases with stellar mass in simulated dwarf galaxies, while \citet{tremmel2024} showed that environment and assembly history also play an important role, with cluster dwarf galaxies being more likely to host massive BHs than field dwarfs. Finally, \citet{bhowmick2025} emphasized that the inferred occupation fraction depends strongly on the adopted BH mass threshold, especially in the regime $M_{\rm BH}\sim 10^5-10^6 \,{\rm M}_{\odot}$, where signatures of BH seeding models may still survive. A recent extension of the BRAHMA simulations further showed that both the seed prescription and the numerical treatment of BH dynamics can strongly affect the predicted occupation fraction \citep{bhowmick2026}.

Semi-analytical models have also been used to investigate the formation and evolution of massive BH populations. \citet{dayal2018}, using the \emph{Delphi} model, followed the hierarchical assembly of galaxies and BHs at high redshift while exploring different BH seed populations and subsequent growth through accretion and mergers. Similarly, \citet{trinca2022} investigated the co-evolution of nuclear BHs and their host galaxies at high redshift, showing that different seeding and accretion scenarios can leave distinct signatures in the low-mass BH population. Using the \emph{L-Galaxies} framework, \citet{spinoso2023} explored multiple BH seed formation channels and showed that the predicted BH occupation fraction can depend strongly on the adopted seeding efficiency. More recently, \citet{bonoli2025} investigated BH seeding and early growth within the \emph{L-GalaxiesBH} framework using constraints from both high-redshift observations and the local BH population. Together, these studies further emphasize that predictions for the massive BH population, particularly in low-mass galaxies and at early cosmic epochs, can retain a strong dependence on the assumptions adopted for BH formation and growth.

On the observational side, \citet{miller2015} used X-ray observations of nearby early-type galaxies to place one of the first constraints on the local occupation fraction, showing that a fully occupied population could not be excluded even in galaxies with $M_{\star}<10^{10}\,M_{\odot}$. More recently, \citet{burke2025} inferred a high local occupation fraction using a Bayesian multiwavelength approach, finding that a large fraction of dwarf galaxies may host central BHs even at $M_{\star}\sim10^{7}-10^{8}\,M_{\odot}$. In contrast, \citet{zou2025}, using archival \textit{Chandra} observations of nearby galaxies, found a strong decline of the occupation fraction toward the dwarf-galaxy regime, with $f_{\rm BH,occ}$ dropping to about one third at $M_{\star}\sim10^{8}-10^{9}\,M_{\odot}$.

A key complication is that observations usually detect only actively accreting BHs. Therefore, the AGN or active fraction should not be directly identified with the intrinsic BH occupation fraction, unless assumptions are made about luminosity limits, duty cycle, accretion-rate distributions, and contamination from stellar processes.

The discussion above highlights an important difficulty in interpreting the BH occupation fraction. On the theoretical side, its predicted shape and normalization can depend on the assumptions adopted for BH formation and growth, while on the observational side the inferred occupation fraction is affected by detection limits and selection effects. In this paper, we do not aim to distinguish between light and heavy seeding scenarios, nor do we attempt to model observational selection effects. Instead, we address a complementary question: how strongly does the intrinsic BH occupation fraction--stellar mass relation predicted within a single semianalytic framework depend on the adopted BH mass threshold, galaxy type and environment, redshift, numerical resolution, simulated volume, and sampled galaxy population? Quantifying these dependencies is important before differences in the occupation fraction can be interpreted primarily in terms of BH seeding or growth physics.

In Section~\ref{sec:model}, we provide a brief description of the AGN feedback prescription implemented in our SAM, the set of simulations used, the model calibration, and the sample selection. We also briefly describe the zoom-in hydrodynamical simulation \texttt{NewCluster}, which is used in our analysis. In Section~\ref{sec:results}, we present and discuss our main results, while in Section~\ref{sec:conclusion} we summarize the main conclusions of this study. Stellar masses are derived assuming a \citet{chabrier2003} initial mass function, and all masses are $h$-corrected.

\section[]{Methods}
\label{sec:model}

In this work, we make use of the semianalytic model \texttt{FEGA25} (\textit{Formation and Evolution of GAlaxies}; \citealt{contini2025,contini2025AGN}, hereafter C25b and C25a) to investigate the evolution of the BH occupation fraction across cosmic time. The model is applied to a set of cosmological dark matter-only simulations, from which halo merger trees are extracted. In addition, part of our analysis is complemented by the high-resolution zoom-in simulation \texttt{NewCluster} \citep{han2026}, which provides useful insight into the evolution of galaxies in dense environments.

Below, we summarize only the aspects of \texttt{FEGA25} that are most relevant for the present work, while referring the reader to C25a and C25b for a complete description of the model and its calibration. In particular, the implementations of supernova (SN) feedback and the AGN-driven hot-gas ejection modes (\texttt{AGNeject1} and \texttt{AGNeject2}) are described in detail in C25a and are therefore not repeated here.

\subsection{\texttt{FEGA25}}

\texttt{FEGA25} is an updated version of the semianalytic framework  presented in \cite{contini2024d} (hereafter C24), including a revised treatment of star formation and AGN feedback. Relative to the previous implementation, the model adopts an extended Kennicutt--Schmidt relation \citep{shi2011} and a more self-consistent coupling between BH growth and AGN activity.

The AGN framework implemented in \texttt{FEGA25} combines different feedback channels operating simultaneously. Along with the standard radio-mode feedback suppressing gas cooling, the model also includes a positive AGN mode, first introduced in C24, capable of triggering additional star formation under specific physical conditions. Furthermore, \texttt{FEGA25} incorporates AGN-driven mechanisms able to expel hot gas beyond the virial radius, thereby regulating the baryonic content of halos. Since these prescriptions are not the main focus of the present paper, we refer the reader to C25a for a detailed discussion of their physical motivation and implementation.

\subsubsection{Quasar Mode}

In \texttt{FEGA25}, BH growth during galaxy mergers follows the classical quasar-mode prescription originally introduced by \cite{croton2006}. During gas-rich mergers, the BH
associated with the central galaxy grows through both the coalescence with the BH of the merging companion and the accretion of part of the available cold gas reservoir.

The gas-accretion contribution to the BH mass increase is modeled as

\begin{equation}
\Delta M_{\rm BH} =
f_{\rm BH}
\left(
\frac{M_{\rm sat}}{M_{\rm cen}}
\right)
\frac{M_{\rm cold}}
{1+\left(280~{\rm km~s^{-1}}/V_{200}\right)^2},
\end{equation}
where $M_{\rm sat}/M_{\rm cen}$ denotes the baryonic mass ratio between the satellite and central galaxies, $M_{\rm cold}$ is the available cold gas mass, and $V_{200}$ is the virial
velocity of the host halo. The parameter $f_{\rm BH }$ regulates the efficiency of gas accretion onto the BH and is fixed to the value adopted in \cite{henriques2020}, i.e., 0.066 \footnote{We keep $f_{\rm BH}$ fixed rather than including it in the MCMC optimization for several reasons. First, this parameter is inherited from the calibration adopted in \citet{henriques2020} and has been retained in the subsequent development of the model. Second, our MCMC calibration is primarily constrained by the evolution of the galaxy stellar mass function, which does not provide a direct constraint on the efficiency of merger-driven BH accretion. Allowing $f_{\rm BH}$ to vary would therefore introduce an additional, relatively poorly constrained degree of freedom and potentially increase degeneracies with other parameters regulating AGN activity. For these reasons, we retain the value adopted in \citet{henriques2020} and calibrate only the parameters that are more directly constrained by the observables used in the MCMC analysis.}.
Following a merger event, the final BH  mass is given by the sum of the preexisting BHs of the progenitor galaxies plus the accreted gas contribution computed above.

We note that, although the detailed relative contribution of the different BH growth modes is not the focus of the present work, the quasar mode still provides an important channel for BH growth,
particularly during merger-driven phases at high redshift.

We emphasize that \texttt{FEGA25} does not assume the presence of preexisting BH seeds. In the model, BHs are formed naturally through gas accretion during quasar-mode events and subsequently evolve through the standard growth channels implemented in the framework, including merger-driven growth and radio-mode accretion (see C24 for detailed analysis on BH mass relations).

We note that, although BH formation in \texttt{FEGA25} is initiated through merger-driven quasar-mode events, subsequent BH growth is not exclusively merger-driven, since radio-mode accretion provides an additional growth channel. The model does not, however, include secular cold-gas accretion triggered by processes such as disc instabilities. The inclusion of such channels could increase BH growth in galaxies with relatively quiet merger histories and may therefore affect the predicted occupation fraction, particularly near the adopted BH mass thresholds.

\subsubsection{Radio Mode and Positive AGN Feedback}

The standard radio-mode AGN feedback follows the prescription introduced by \cite{croton2006}. In this framework, hot halo gas accretes onto the central supermassive BH, releasing mechanical energy that partially or fully suppresses radiative cooling.

The BH accretion rate is modeled as

\begin{equation}
\dot{M}_{\rm BH} =
\kappa_{\rm AGN}
\left(\frac{f_{\rm hot}}{0.1}\right)
\left(\frac{V_{200}}{200~{\rm km~s^{-1}}}\right)^3
\left(\frac{M_{\rm BH}}{10^8~M_\odot h^{-1}}\right),
\end{equation}
where $f_{\rm hot}$ is the hot-gas fraction within the halo and $\kappa_{\rm AGN}$ is a free parameter calibrated through MCMC techniques.

The energy released by AGN accretion is

\begin{equation}
E_{\rm radio} =
\eta_{\rm rad} \dot{M}_{\rm BH} c^2 ,
\end{equation}
with $\eta_{\rm rad}=0.1$.
The cooling rate is then modified according to

\begin{equation}
\dot{M}_{\rm cool,new} =
\dot{M}_{\rm cool}
-
\frac{2E_{\rm radio}}{V_{200}^2}.
\end{equation}

Whenever cooling is not completely suppressed, \texttt{FEGA25} allows AGN activity to induce an additional contribution to star formation through a positive feedback mode. This prescription, originally introduced in C24 and further revised in C25b, links the efficiency of the positive mode directly to BH accretion, avoiding the introduction of additional free parameters.

The induced star formation is written as

\begin{equation}
\Delta M_{*,{\rm AGN}} =
\frac{\Delta M_{\rm BH}}
{M_{\rm BH}}
\frac{\dot{M}_{\rm cool,new}}
{10^8~M_\odot h^{-1}} .
\end{equation}

This formulation naturally couples positive and negative AGN feedback. At high redshift, when cooling is still efficient, the positive mode may contribute significantly to star formation.
At later epochs, as AGN heating becomes progressively more effective, the positive contribution weakens.

\subsubsection{Set of Simulations and Calibration}

The calibration of \texttt{FEGA25} follows the same MCMC strategy presented in C24 and C25a/C25b. Model parameters are constrained by reproducing the observed evolution of the stellar mass function from $z=3$ to $z=0$.

The observational compilation includes measurements from \cite{marchesini2009,marchesini2010}, \cite{ilbert2010,ilbert2013}, \cite{muzzin2013}, \cite{tomczak2014}, \cite{baldry2008,baldry2012}, \cite{li-white2009}, and \cite{bernardi2018}.

Galaxy catalogs are generated using merger trees extracted from three cosmological dark matter-only simulations: YS50HR (hereafter YS50), YS200, and YS300, corresponding to box sizes of $50$, $200$, and $300~{\rm Mpc}~h^{-1}$, respectively. All simulations were run with GADGET4 \citep{springel2021} assuming the Planck 2018 cosmology \citep{planck2020}.
The corresponding dark matter particle masses are $10^7$, $3.26\times10^8$, and $2.2\times10^9~M_\odot h^{-1}$.

Following C25a, the main free parameters relevant for the present work are the AGN accretion efficiency $\kappa_{\rm AGN}$, the characteristic velocity scale regulating AGN-driven gas ejection, and the reincorporation parameter $\gamma_2$. The best-fit values adopted in this paper are the same as those presented in C25a. To construct the final galaxy catalogs, \texttt{FEGA25} is applied to all three simulation sets. For a detailed description of the numerical setup, time sampling, resolution properties, and validation of the three simulations, we refer the reader to C25a and references therein.

\subsection{\texttt{NewCluster}}

The \texttt{NewCluster} simulation \citep{han2026} is a cosmological zoom-in simulation designed to study galaxy formation within a massive cluster environment. The simulation follows the evolution of a galaxy cluster with virial mass $\sim5\times10^{14}~M_\odot$ at $z=0$.

The run was performed with the RAMSES-yOMP code \citep{han2025a}, a hybrid-parallel version of the RAMSES adaptive mesh refinement hydrodynamics solver \citep{teyssier2002}. The parent simulation box has a size of $100~{\rm Mpc}~h^{-1}$, while the zoomed region extends to approximately $3.5R_{\rm vir}$ ($\sim 20 {\rm Mpc/h}$) around the central cluster. Within the high-resolution region, the dark matter particle mass is $m_{\rm DM}=1.3\times10^6~M_\odot$, while stellar particles have mass $m_*=2\times10^4~M_\odot$. The best spatial resolution is approximately $70$ pc in comoving units in the high-resolution region.

The simulation includes radiative cooling, star formation, chemical enrichment, SN feedback, and AGN feedback. Gas cooling accounts for both metal-line and dust cooling, while star formation follows a thermoturbulent efficiency prescription in dense gas regions.

AGN feedback adopts the dual-mode scheme proposed by \cite{dubois2012}, with kinetic jet feedback operating at low Eddington ratios and thermal quasar-mode feedback active at high accretion rates. \texttt{NewCluster} adopts a BH seed mass of $10^5\,M_\odot$; for a detailed description of the full seeding prescription, we refer the reader to \cite{han2026}.

Although \texttt{NewCluster} is not the primary focus of the present paper, it provides a useful comparison for assessing the baryonic and BH properties predicted by \texttt{FEGA25},
especially at high resolution and in dense environments. We stress that NewCluster samples a highly overdense region and is therefore not expected to be representative of an average cosmological volume. Its comparison with \texttt{FEGA25} should consequently be regarded as qualitative, since differences in the predicted BH occupation fraction may reflect not only the different numerical and physical modelling, but also the particular environment and assembly history sampled by the zoom-in simulation.

\subsection{Sample Selection and Resolution Limits}

To ensure adequate numerical resolution, we construct our galaxy sample by applying stellar-mass and halo-mass selection criteria. At all redshifts considered, galaxies are required to satisfy $\log M_{\star}>7$. In addition, a minimum halo-mass threshold is imposed in each simulation. These limits correspond approximately to halos resolved with $\sim10^4$ dark matter particles. The adopted thresholds
are

\[
\log M_{\rm halo,min} =
\left\{
\begin{array}{ll}
11.0  & \mathrm{YS50} \\
12.51 & \mathrm{YS200} \\
13.35 & \mathrm{YS300}
\end{array}
\right.
\]

The same halo-mass thresholds are applied at all redshifts considered. This conservative selection is designed to minimize numerical effects associated with poorly resolved halos. This is particularly important for satellite galaxies, whose subhalos can be strongly affected by tidal stripping and may be lost or disrupted in the merger trees, especially at low masses.

\section{Results and Discussion}
\label{sec:results}

In this section, we present the results of our analysis, focusing on the relation between the BH occupation fraction, $f_{\rm BH,occ}$, and galaxy stellar mass. We investigate how this relation depends on the adopted BH mass threshold, galaxy type and environment, simulated volume, numerical resolution, sampled galaxy population, and redshift. Finally, we compare the predictions of our model with both observational constraints and results from other numerical methods.

\begin{figure*}[t!]
\centering
\includegraphics[width=0.94\textwidth]{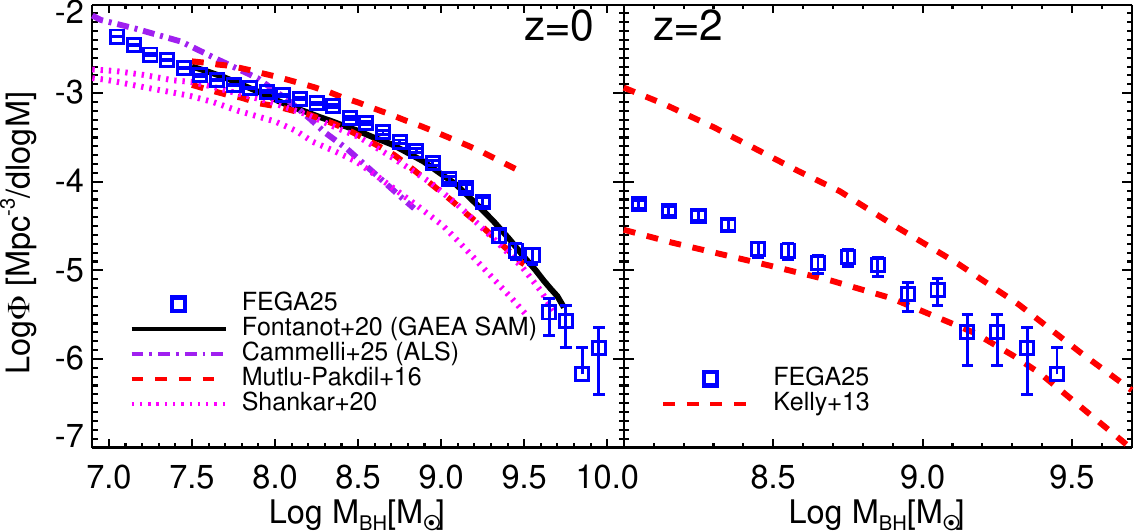}
\caption{Left panel: Black hole mass function at $z=0$ predicted by our model (blue squares), compared with the predictions of the \texttt{GAEA} semianalytic model (\citealt{fontanot2020}; black solid line) and its All Light Seed implementation (\citealt{cammelli2025}; purple dash-dotted line). Observational constraints from \citet{mutlu2016} and \citet{shankar2020} are
shown as red dashed and magenta dotted lines, respectively. Right panel: Same as the left panel, but at $z=2$, where our predictions are compared with the observational estimate of \citet{kelly2013} (red dashed line). All observational curves show the corresponding $\pm \sigma$ scatter. At $z=0$, the predicted black hole mass function is in good agreement
with both the observational constraints and the theoretical predictions, while at $z=2$ it is consistent with the observational estimate of \citet{kelly2013}. The \texttt{FEGA25} predictions shown in both panels are obtained from the YS200 simulation.}
\label{fig:fig1}
\end{figure*}

Before addressing these points, we first verify that our SAM is able to reproduce the observed distribution of BH masses, particularly in the local Universe. Figure~\ref{fig:fig1} shows the BH mass function at $z=0$ (left panel) and $z=2$ (right panel). The predictions of \texttt{FEGA25} are shown as blue squares. At $z=0$, they are compared with a set of observational constraints and with predictions from the \texttt{GAEA SAM} \citep{fontanot2020}, as indicated in the legend. At $z=2$, we compare our model with the observational estimate of \citet{kelly2013}.

Our predictions are in good agreement with previous observational \citep{kelly2013,mutlu2016,shankar2020} and theoretical \citep{fontanot2020,cammelli2025} results. We stress, however, that the comparison with the theoretical predictions is intended to place the \texttt{FEGA25} results in the context of other galaxy-formation models, rather than to provide an independent validation of the model. The agreement with the observational constraints provides the primary consistency check for the following analysis, as it shows that the model reproduces the global distribution of BH masses from $z=2$ to the present day.

For completeness, we compare in Appendix~\ref{app:bhmf} the BH mass functions predicted from YS50, YS200, and YS300 at $z=0$, $1$, $2$, and $3$, in order to illustrate the simulation-to-simulation variations associated with the different volumes and mass resolutions, and sampled galaxy populations. We therefore now turn to the main focus of this work, namely the dependence of the $f_{\rm BH,occ}-M_{\rm gal}$ relation on the physical and numerical factors outlined above.

\subsection{Dependence on Galaxy Type and BH Mass Threshold}
\label{sec:environment}

We first investigate how the predicted $f_{\rm BH,occ}$ depends on galaxy type and on the adopted BH mass threshold. The distinction between central and satellite galaxies is important
because the two populations experience different evolutionary paths. Central galaxies can continue to accrete gas when available, while satellite galaxies may be affected by environmental processes such as
ram-pressure stripping \citep{gunngott1972}, starvation \citep{larson1980}, and tidal stripping \citep[e.g.,][]{contini2014, contini2024a}. These mechanisms can reduce the gas reservoir available for subsequent star formation and BH growth. Their impact is expected to depend not only on stellar mass, but also on the time since infall and on the environment into which the satellite is accreted.

Before starting the analysis, we specify how the error bars are estimated. We assume binomial statistics. In each stellar-mass bin, the fraction $f$ is computed as the ratio between the number of objects satisfying the considered criterion and the total number of objects in the bin. The corresponding uncertainty is given by
\[
\sigma_f = \sqrt{\frac{f(1-f)}{N}},
\]
where $N$ is the denominator used to compute the fraction. The error bars show $f\pm\sigma_f$.

\begin{figure*}[t!]
\centering
\includegraphics[width=0.95\textwidth]{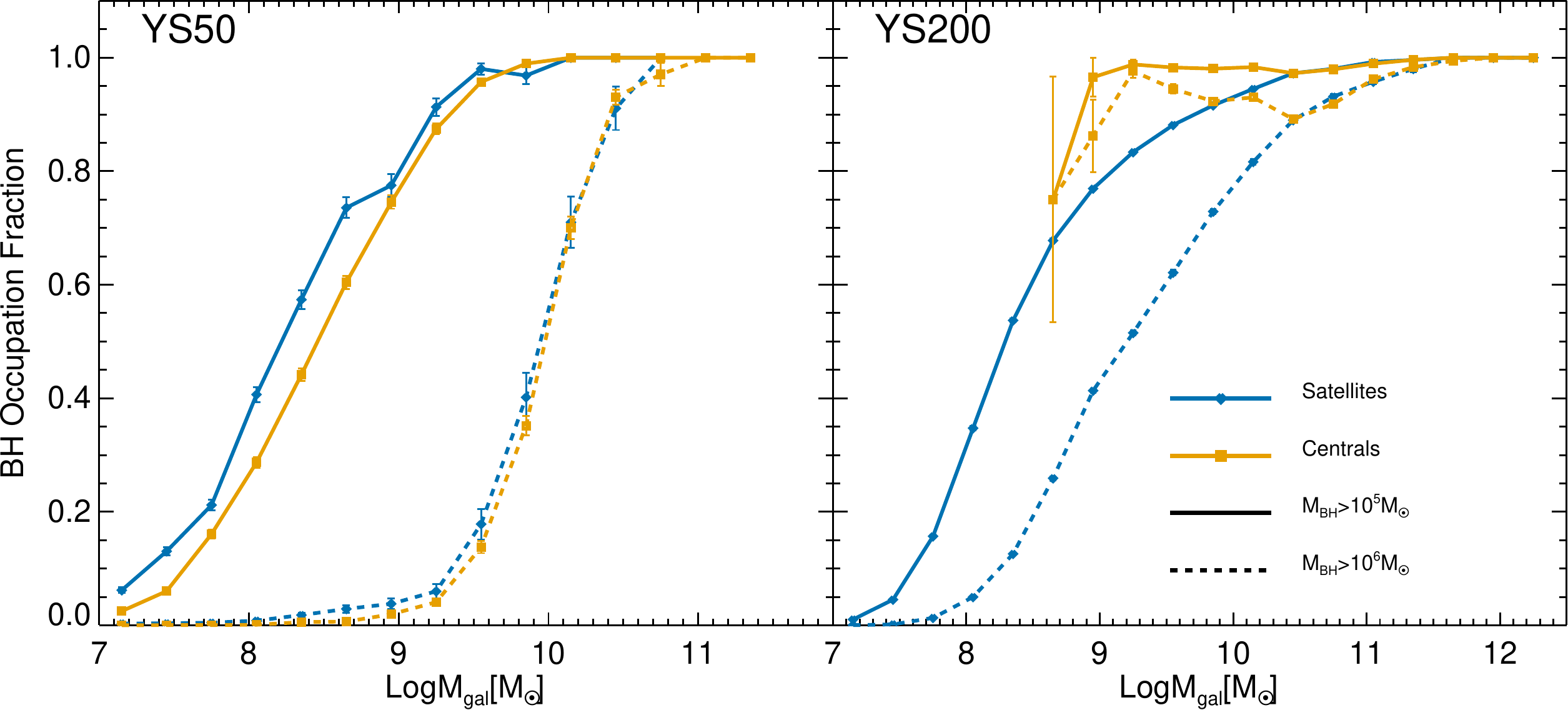}
\caption{Black hole occupation fraction, $f_{\rm BH,occ}$, as a function of stellar mass at $z=0$. The left and right panels show YS50 and YS200, respectively. Colors and symbols distinguish satellite galaxies from central galaxies, while line styles indicate the adopted black hole mass threshold: solid lines for $M_{\rm BH}>10^5\,M_\odot$ and dashed lines for $M_{\rm BH}>10^6\,M_\odot$. The comparison shows that the relative behavior of central and satellite galaxies depends on both the simulation box and the adopted black hole mass threshold. In YS50,
satellites lie slightly above centrals for the lower threshold, while the two populations show very similar relations for the higher threshold. In YS200, central galaxies reach high occupation fractions at lower stellar masses than satellites. At the high-mass end, the differences between the two populations become weaker, as most galaxies host black holes above the adopted threshold.}
\label{fig:fig2}
\end{figure*}

Figure~\ref{fig:fig2} shows $f_{\rm BH,occ}$ as a function of stellar mass for central and satellite galaxies in YS50 and YS200 at $z=0$. The two panels correspond to the two simulation boxes, while colors separate central and satellite galaxies. Different line styles indicate the two adopted BH mass thresholds, $M_{\rm BH}>10^5 \,{\rm M}_{\odot}$ and $M_{\rm BH}>10^6 \,{\rm M}_{\odot}$.

In both simulations, the occupation fraction increases with stellar mass. This is expected, since more massive galaxies generally experience more efficient BH growth through gas accretion and mergers. However, the stellar mass at which the occupation fraction approaches unity depends on both galaxy type and BH mass threshold. For the lower threshold, $M_{\rm BH}>10^5 \,{\rm M}_{\odot}$, the transition to high occupation fractions occurs at lower stellar masses than for the higher threshold, $M_{\rm BH}>10^6 \,{\rm M}_{\odot}$. This shift is a direct consequence of applying a more restrictive BH mass selection.

The comparison between central and satellite galaxies shows that their relative behavior is not universal, but depends on both the simulation box and the adopted BH mass threshold. In YS50, satellites have slightly higher occupation fractions than centrals over much of the transition regime for the lower threshold, whereas the two populations show very similar relations for the higher threshold. In YS200, by contrast, central galaxies already show high occupation fractions over most of the sampled stellar-mass range, while satellites exhibit a smoother increase with stellar mass. The central--satellite difference is therefore box-dependent and is most relevant in the transition regime, where BH masses are close to the adopted threshold. At the high-mass end, most
galaxies already host BHs above the adopted threshold, and the distinction between centrals and satellites becomes less important for $f_{\rm BH,occ}$.

The BH mass threshold has a particularly strong effect on the satellite population. Satellite galaxies are more likely to host BHs with masses close to the adopted threshold, so increasing the BH mass threshold from $10^5$ to $10^6 \,{\rm M}_{\odot}$ shifts the satellite occupation fraction toward higher stellar masses. In contrast, massive central galaxies often host BHs above both thresholds, making their occupation fraction less sensitive to the exact value of the BH mass threshold once sufficiently large stellar masses are reached.

\begin{figure*}[t!]
\centering
\includegraphics[width=0.94\textwidth]{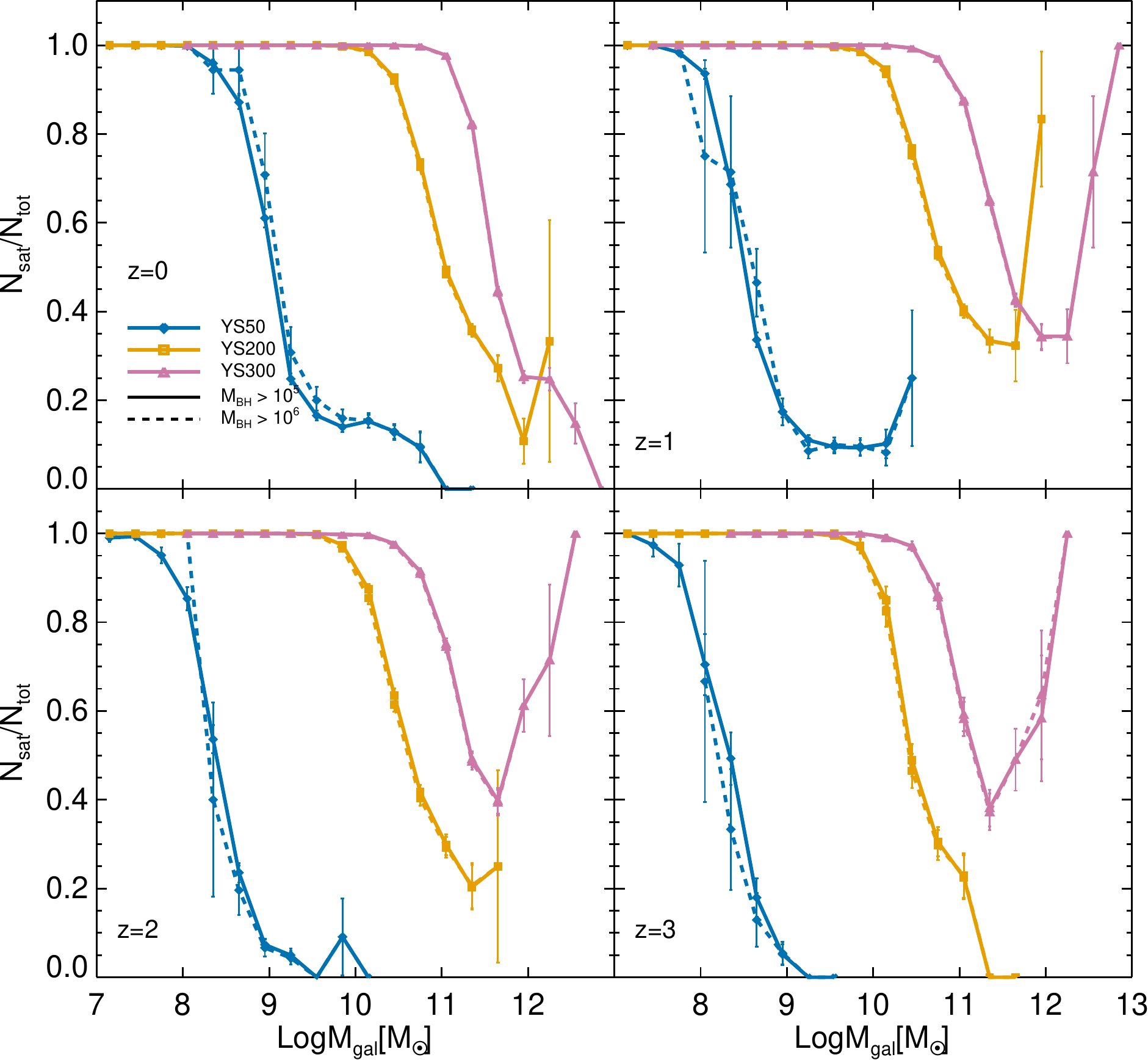}
\caption{Fraction of satellite galaxies as a function of stellar mass for our sample. The four panels correspond to $z=0$, $z=1$, $z=2$, and $z=3$, as indicated. In each stellar-mass bin, the plotted quantity is defined as $N_{\rm sat}/N_{\rm tot}$, where $N_{\rm sat}$ is the number of satellite galaxies hosting a black hole above the adopted mass threshold, and $N_{\rm tot}$ is the total number of galaxies, centrals plus satellites, satisfying the same black hole mass selection. Colors and symbols identify the simulation box: YS50 is shown in blue with diamonds, YS200 in orange with squares, and YS300 in purple with triangles, while line styles indicate the two black hole mass thresholds, with solid lines corresponding
to $M_{\rm BH}>10^5\,{\rm M}_{\odot}$ and dashed lines to $M_{\rm BH}>10^6\,{\rm M}_{\odot}$. The figure shows that the relative contribution of satellites depends strongly on stellar mass, redshift, and simulation box. In particular, the satellite contribution is high at the low-mass end in the selected samples, while it decreases toward higher stellar masses, with the transition occurring at different masses in the three simulation boxes. This confirms that the total black hole occupation fraction should be interpreted as a population-weighted quantity, whose shape
can be significantly affected by the relative abundance of central and satellite galaxies in each stellar-mass bin.}
\label{fig:fig3}
\end{figure*}

The occupation fraction measured for the full galaxy population should therefore be interpreted as a population-weighted quantity. It depends not only on the occupation fractions of central and satellite galaxies
separately, but also on their relative abundance in each stellar-mass bin. To quantify this point, Figure~\ref{fig:fig3} shows $N_{\rm sat}/N_{\rm tot}$ as a function of stellar mass, where $N_{\rm sat}$ is the number of satellite galaxies hosting a BH above the adopted threshold and $N_{\rm tot}$ is the total number of BH-hosting galaxies above the same threshold in the bin.

Figure~\ref{fig:fig3} shows that the satellite contribution depends strongly on stellar mass, redshift, and simulation box. In the selected BH-hosting population, satellites provide a large contribution
at the low-mass end, while their relative importance generally decreases toward higher stellar masses. The stellar mass at which this decrease occurs shifts systematically among the simulations, moving to higher
masses from YS50 to YS300. This reflects the fact that the three boxes sample different galaxy populations, with different mass resolution, volume, and relative abundance of central and satellite galaxies.

As an additional test, we computed $N_{\rm sat}/N_{\rm tot}$ using the full galaxy population, without imposing any lower BH mass threshold. The differences among YS50, YS200, and YS300 remain essentially unchanged, indicating that the simulation-to-simulation variations are primarily driven by differences in the underlying central--satellite galaxy populations rather
than by the adopted BH mass selection.

In some cases, however, $N_{\rm sat}/N_{\rm tot}$ increases again at the high-mass end. We do not interpret this upturn as direct evidence for enhanced BH growth in satellites. At these stellar masses, the
number of objects per bin is small, and the ratio can therefore be sensitive to a few massive satellites. Moreover, massive satellites are often systems that were accreted after having already grown a BH
above the adopted threshold. The high-mass upturn therefore mainly reflects the demographic composition of the selected BH-hosting population, rather than a simple environmental effect.

This result clarifies the interpretation of the total occupation fraction. When satellites represent a substantial fraction of the BH-hosting population in a given stellar-mass bin, the total $f_{\rm BH,occ}$ can closely follow the satellite contribution. When centrals dominate, the total relation is instead expected to be closer to the central-galaxy occupation fraction. Thus, the global $f_{\rm BH,occ}-M_{\star}$ relation should not be interpreted as a simple universal curve, but as the result of the combined contributions of central and satellite galaxies, weighted by their relative abundance in the selected sample.

Our results are broadly consistent with previous studies showing that $f_{\rm BH,occ}$ increases with stellar mass, including the dwarf-galaxy regime \citep[e.g.,][]{bellovary2019,bhowmick2025}. They are also in qualitative agreement with the idea that environment and assembly history can affect the probability of hosting a massive BH. For example, \citet{tremmel2024} found that cluster dwarf galaxies are more likely to host massive BHs than field dwarfs, and that early-forming systems have a higher probability of growing a BH. Although our analysis does not explicitly follow formation time, the differences between central and satellite galaxies, and the variation of $N_{\rm sat}/N_{\rm tot}$ among the simulation boxes, suggest that the galaxy population sampled in each stellar-mass bin is important for interpreting the total occupation fraction.

The dependence on the adopted BH mass threshold is also in line with the conclusions of \citet{bhowmick2025}, who emphasized that the inferred occupation fraction can change significantly with the chosen
threshold, especially in the regime $M_{\rm BH}\sim 10^5-10^6 \,{\rm M}_{\odot}$. In our model, this effect is particularly visible for satellite galaxies and near the transition regime, where many BHs lie close to the adopted threshold. This also connects to observational studies, where the inferred presence of a BH depends on detection limits and selection effects \citep[e.g.,][]{burke2025}. We return to the comparison with observational constraints and other numerical predictions in Section~\ref{sec:comparison}.

These results show that galaxy type and BH mass threshold are both essential for interpreting the predicted occupation fraction. In the following section, we extend the comparison to the three simulation
boxes in order to assess how the predicted relation changes with simulation volume, mass resolution, and the galaxy population sampled by each box.

\subsection{Dependence on Simulation Volume, Resolution, and Sampled Population}
\label{sec:resolution}

\begin{figure*}[t!]
\centering
\includegraphics[width=0.94\textwidth]{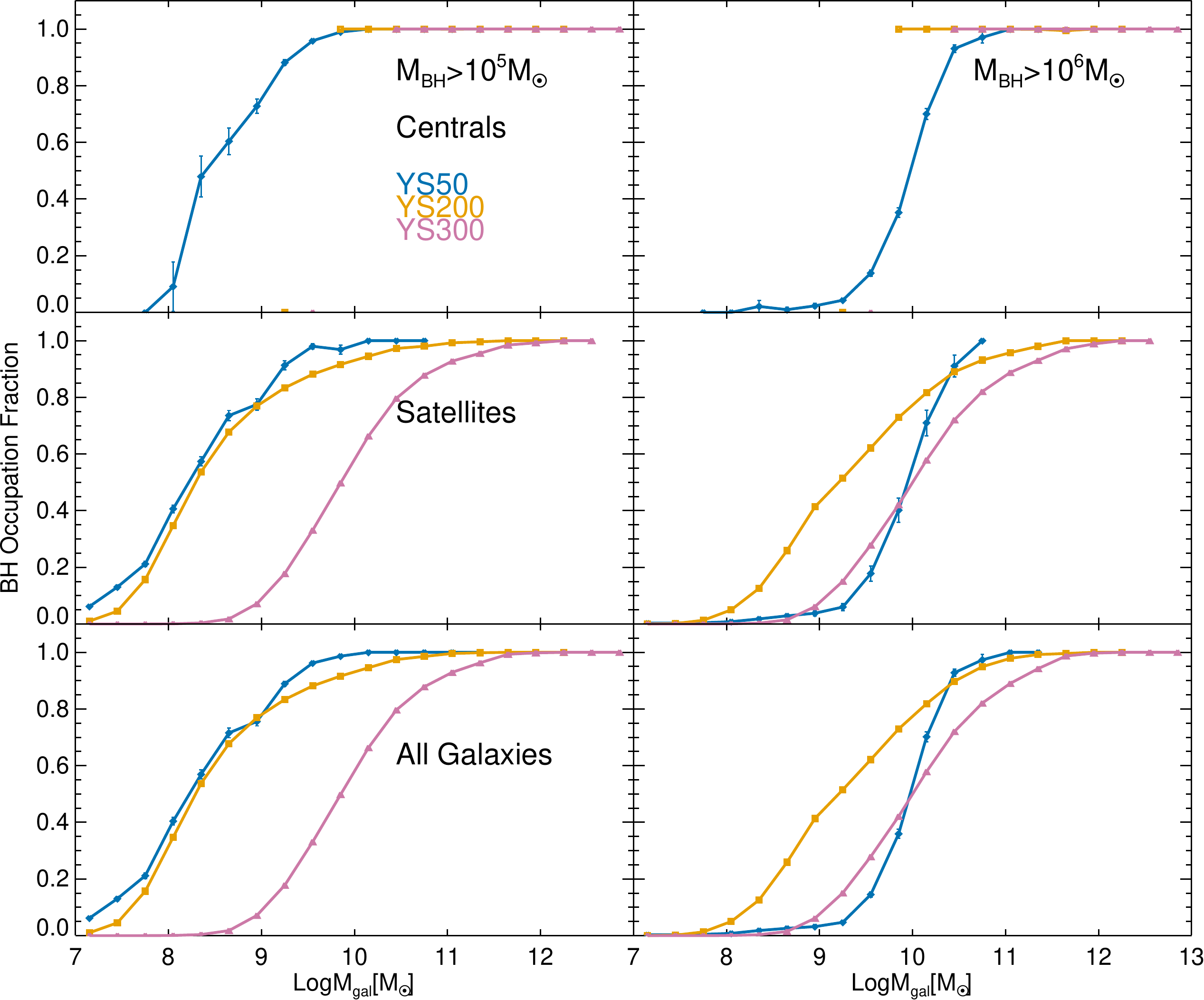}
\caption{Comparison of the black hole occupation fraction ($f_{\rm BH,occ}$) as a function of stellar mass in the three simulation boxes, separately for central galaxies (upper panels), satellite galaxies (middle panels), and the full galaxy population (lower panels). The left column shows the results obtained with the lower black hole mass threshold, while the right column corresponds to the higher threshold. Colors and symbols identify the three simulation boxes: YS50 is shown in blue with diamonds, YS200 in orange with squares, and YS300 in
purple with triangles. Since the simulations span a wide range of mass resolutions, from the highest-resolution box (YS50) to the lowest-resolution one (YS300), these panels illustrate the box-to-box variation of the predicted occupation fraction, reflecting the combined impact of mass resolution, simulated volume, and sampled galaxy population.}
\label{fig:fig4}
\end{figure*}

Having quantified the dependence of $f_{\rm BH,occ}$ on galaxy type and BH mass threshold, we now focus on the simulation-to-simulation variation of the occupation fraction itself. Figure~\ref{fig:fig4} compares the $f_{\rm BH,occ}-M_{\star}$ relation predicted by YS50, YS200, and YS300 for central galaxies, satellites, and the full galaxy population. The three
simulations differ substantially in mass resolution, simulated volume, and the galaxy populations that they sample, allowing us to assess the combined impact of these factors on the predicted occupation fraction.

In the upper panels, YS200 and YS300 show very similar relations for central galaxies, with only a weak dependence on the adopted BH mass threshold over the stellar-mass range sampled by the two simulations.
This indicates that the central galaxies selected in YS200 and YS300 are already massive enough to host BHs above the adopted thresholds with high efficiency. A different behavior, already discussed above, is found for YS50, where the occupation fraction increases more gradually with stellar mass also for central galaxies. This likely reflects the different central-galaxy population sampled by the smaller box, which includes a larger fraction of lower-mass centrals associated with less massive distinct host halos.

The comparison between the three simulations highlights significant box-to-box variations, which reflect the combined effect of mass resolution, simulated volume, and the different galaxy populations
sampled by each box. Let us first focus on the lower BH mass threshold for satellites and for all galaxies combined. The YS50 and YS200 predictions are very close to each other, indicating that, when the adopted threshold is relatively low ($M_{\rm BH}>10^5\,M_{\odot}$, a commonly adopted seed mass in several models; e.g., \citealt{dubois2014,schaye2015,pillepich2018,han2026}), satellites in YS50 can also retain sufficiently massive BHs. In YS300, instead, the relation is significantly lower at the low-mass end, and the occupation fraction reaches $f_{\rm BH,occ}\sim0.5$ only at $M_{\rm gal}\sim10^{10}\,M_{\odot}$, roughly two orders of magnitude higher than in YS50 and YS200. When the BH mass threshold is increased, the relations become more similar in some stellar-mass ranges, although significant box-to-box differences remain.

The systematic differences among the three simulation boxes suggest that the inferred BH occupation fraction depends on a combination of numerical resolution, simulated volume, and the resulting mix of galaxy
environments. The highest-resolution simulation, YS50, resolves lower-mass halos and their associated BHs more effectively, while the lower-resolution boxes progressively miss part of this population, shifting the transition to high occupation fractions toward larger stellar masses. At the same time, the larger volumes sample a wider range of environments, including a larger number of groups and clusters, thereby changing the relative contribution of central and satellite galaxies. Since the total occupation fraction is a population-weighted quantity, differences in the abundance and evolutionary histories of satellites can significantly affect the inferred global relation.

A similar sensitivity to numerical resolution has been discussed in several theoretical studies, where the occupation fraction at the low-mass end depends critically on the ability to resolve both low-mass galaxies and their central BHs (e.g., \citealt{haidar2022,bellovary2019,tremmel2024,bhowmick2025}). These studies consistently show that limited resolution can shift the transition to high occupation fractions toward larger stellar masses and suppress the inferred abundance of BHs in dwarf galaxies. Recent BRAHMA results also show that the predicted occupation fraction can depend on the numerical treatment of BH dynamics:
\citet{bhowmick2026} find that BH repositioning can artificially suppress the occupation fraction at low stellar masses and low BH mass thresholds, whereas a subgrid dynamical-friction treatment allows more low-mass
galaxies to retain their BHs.

\subsection{Dependence on Redshift}
\label{sec:redshift}

The redshift evolution of $f_{\rm BH,occ}$ provides a direct way to investigate how BH growth and galaxy assembly proceed across cosmic time. At fixed stellar mass, changes in $f_{\rm BH,occ}$ may reflect several effects acting simultaneously: the early formation or growth of BHs, the subsequent build-up of the galaxy population, environmental processing, and the changing relative contribution of central and satellite galaxies. In particular, galaxies with the same stellar mass at different redshifts do not necessarily represent the same evolutionary population, since they may differ in formation time, merger history, gas availability, and environment.

For this reason, the redshift evolution of $f_{\rm BH,occ}$ should not be interpreted as a simple function of stellar mass alone. Moreover, direct observational constraints remain limited, because observations generally detect only actively accreting BHs above a luminosity threshold, whereas theoretical models can track the intrinsic $f_{\rm BH,occ}$ above a chosen BH mass threshold. Inferring the redshift dependence of the intrinsic occupation fraction from observations is therefore not trivial, as it requires assumptions about duty cycle, accretion-rate distributions, luminosity limits, and selection effects (e.g., \citealt{greene2012,burke2025,zou2025}). In what follows, we therefore examine how $f_{\rm BH,occ}$ evolves from $z=3$ to the present day in our different simulation boxes, and assess whether the inferred trends depend on the simulated volume and galaxy population sampled.

\begin{figure*}[t!]
\centering
\includegraphics[width=0.94\textwidth]{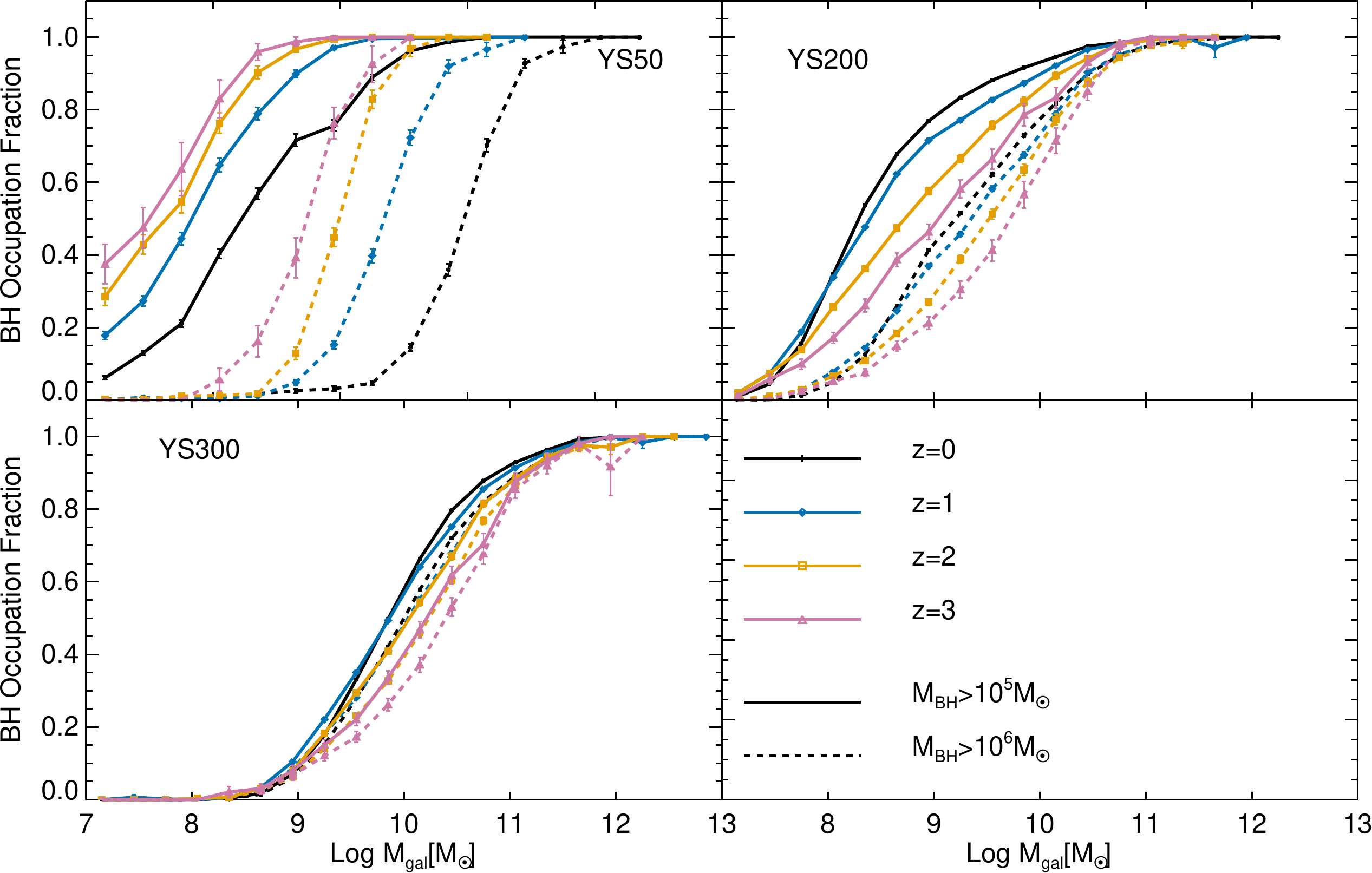}
\caption{Redshift dependence of the black hole occupation fraction ($f_{\rm BH,occ}$) as a function of stellar mass from $z=3$ to the present day. Colors and symbols identify the different redshifts: $z=0$ is shown in black with plus signs, $z=1$ in blue with diamonds, $z=2$ in orange with squares, and $z=3$ in purple with triangles. Each panel corresponds to one of the three simulation boxes and includes the lower (solid lines) and higher (dashed lines) black hole mass thresholds. In YS50, the occupation fraction decreases toward lower redshift, whereas
the opposite trend is found in YS200 and YS300, with the effect being more pronounced in YS200 and largely independent of the adopted threshold. This behavior may reflect both the different environments sampled by the three simulation volumes and the fact that galaxies with the same stellar mass at different redshifts can be in different evolutionary stages.}
\label{fig:fig5}
\end{figure*}

Figure~\ref{fig:fig5} addresses these points by showing the redshift dependence of the $f_{\rm BH,occ}-M_{\rm gal}$ relation in YS50, YS200 and YS300 (different panels), from $z=3$ to the present time (different colors as indicated in the legend), and for the two BH mass thresholds (different line styles). We see that the evolutionary trend is rather different among the three boxes. In the smallest box, the relation in galaxies with mass lower than around $M_{\rm gal}=10^{9.5} M_{\odot}$ decreases toward lower redshift, while the opposite trend is found in the other two boxes, most clearly in YS200, regardless of the BH  mass threshold.

The different trends may reflect, at least in part, the different environments and galaxy populations sampled by the three simulation boxes. Moreover, galaxies occupying the same stellar-mass bin at different redshifts do not necessarily correspond to the same evolutionary population. Therefore, the evolution of $f_{\rm BH,occ}$ measured at fixed stellar mass should not be
interpreted as directly tracing the evolution of individual galaxies or their descendants. Establishing the role of progenitor bias quantitatively would require tracking galaxy and BH assembly histories across cosmic time, which is beyond the scope of the present analysis.

\begin{figure*}[t!]
\centering
\includegraphics[width=0.94\textwidth]{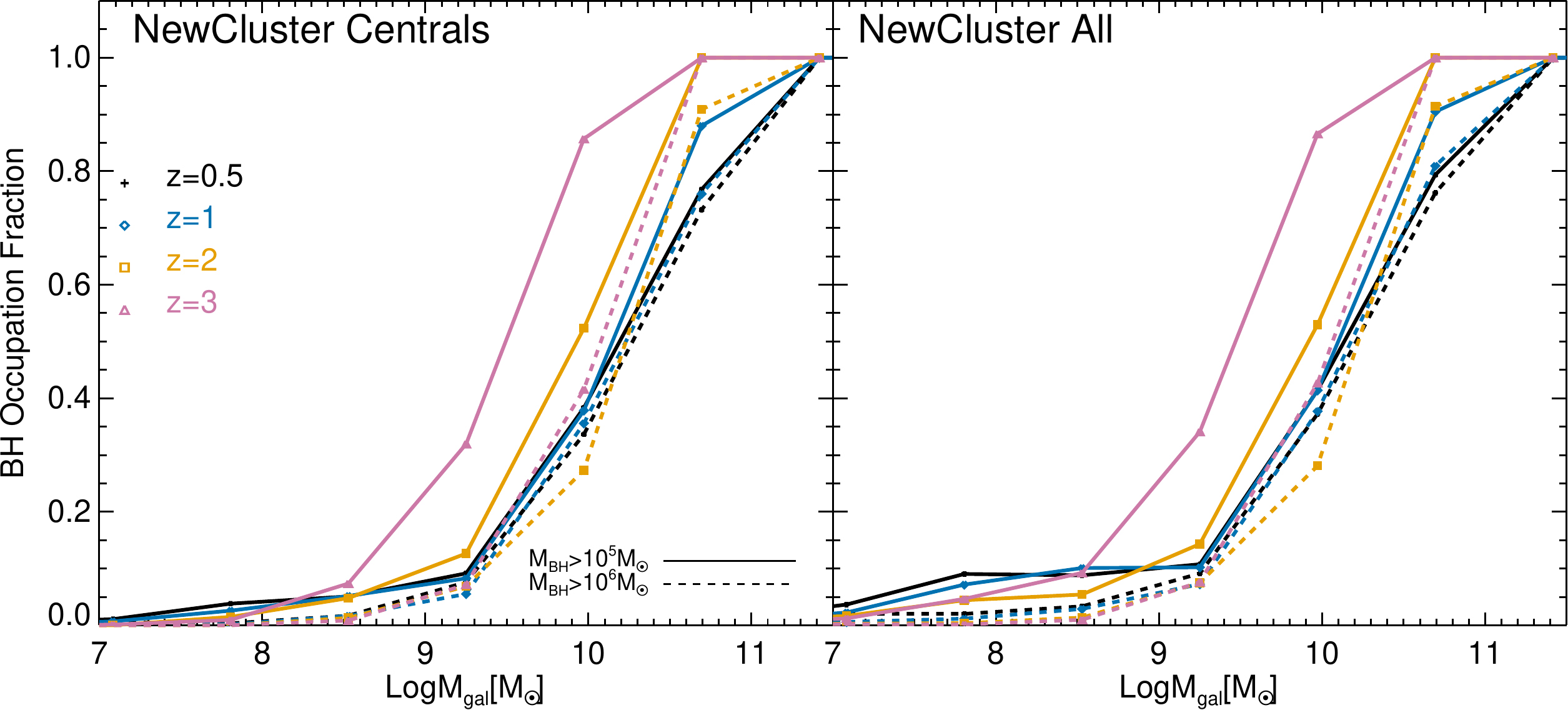}
\caption{Same as Figure~\ref{fig:fig5}, but using data from our in-house \textit{NewCluster} simulation. Galaxies are separated into central galaxies only (left panel) and the full population including satellites (right panel). The same black hole mass thresholds adopted in the previous figures are used. NewCluster shows a redshift evolution qualitatively similar to that found in YS50 and to the behavior reported by \citet{tremmel2024} for Romulus25 and RomulusC.}
\label{fig:fig6}
\end{figure*}

So far we have predictions only for cosmological boxes that sample different environments, and it would be interesting to know what zoom-in simulations predict. For this purpose, we make use of \texttt{NewCluster} for the first time, and plot in Figure~\ref{fig:fig6} the evolution in redshift of the predicted relation, from $z=3$ to $z=0.5$ (\texttt{NewCluster} is still ongoing and has not reached $z=0$ yet), for centrals only in the left panel and all galaxies in the right one.
\texttt{NewCluster} provides an additional qualitative comparison and shows a redshift evolution broadly similar to that found in YS50, for galaxies with $M_{\rm gal}>10^9M_{\odot}$. We stress, however, that \texttt{NewCluster} samples a highly overdense region associated with the progenitor of a massive cluster, and therefore its galaxy population is not expected to be
representative of an average cosmological environment. Part of the difference between \texttt{NewCluster} and the cosmological boxes may consequently arise from environmental and assembly-history effects, rather than solely from differences in the numerical or physical modelling. For this reason, the comparison should be regarded as qualitative.

Overall, our results indicate that the redshift evolution of $f_{\rm BH,occ}$ cannot be interpreted as a universal trend at fixed stellar mass. While YS50 predicts a decreasing occupation fraction toward lower redshift, in qualitative agreement with the behavior found by \citet{tremmel2024} in Romulus25, the larger-volume boxes show the opposite trend. A qualitatively similar redshift trend has recently been reported by \citet{bhowmick2026}, whose BRAHMA simulations predict higher occupation fractions at fixed stellar mass at $z\sim5$ than at $z=0$, particularly for the lower BH mass thresholds.
The qualitative similarity between YS50 and \texttt{NewCluster} is reminiscent of the parallel between Romulus25 and RomulusC reported by \citet{tremmel2024}, although the comparison should not be interpreted as one-to-one because of the different numerical methods, resolutions, and galaxy selections. This suggests that the evolution of $f_{\rm BH,occ}$ is not controlled by stellar mass alone, but reflects the combined effect of BH growth, galaxy assembly history, environment, numerical resolution, sampled galaxy population, and the changing mix of central and satellite galaxies across cosmic time. Direct theoretical comparisons remain limited, as only a few studies explicitly follow $f_{\rm BH,occ}$ as a function of both stellar mass and redshift. Therefore, the different trends found among our simulations should be interpreted as evidence that the redshift evolution of $f_{\rm BH,occ}$ is sensitive to the galaxy population sampled at each epoch, rather than as a single universal prediction.

On the other hand, direct observational constraints on the redshift evolution of $f_{\rm BH,occ}$ remain extremely limited. Current studies provide robust estimates primarily in the local Universe (e.g., \citealt{burke2025,zou2025}), while at higher redshift most measurements are restricted to the fraction of actively accreting BHs, which depends strongly on duty cycle and selection effects (e.g., \citealt{mezcua2018}). A direct comparison between intrinsic occupation fractions and observed active fractions would therefore require modelling the duty cycle, luminosity limits, and selection functions of the surveys.

\subsection{Comparison with Other Studies}
\label{sec:comparison}

\begin{figure*}[t!]
\centering
\includegraphics[width=0.95\textwidth]{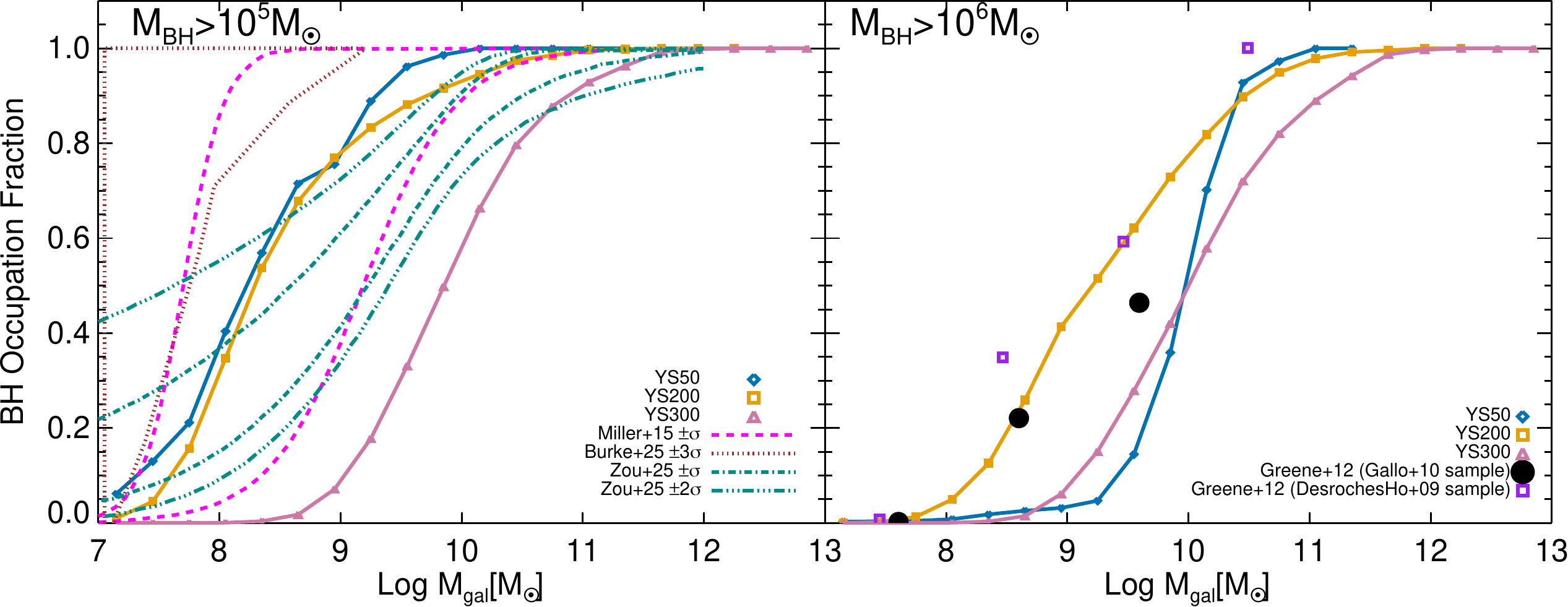}\\
\includegraphics[width=0.95\textwidth]{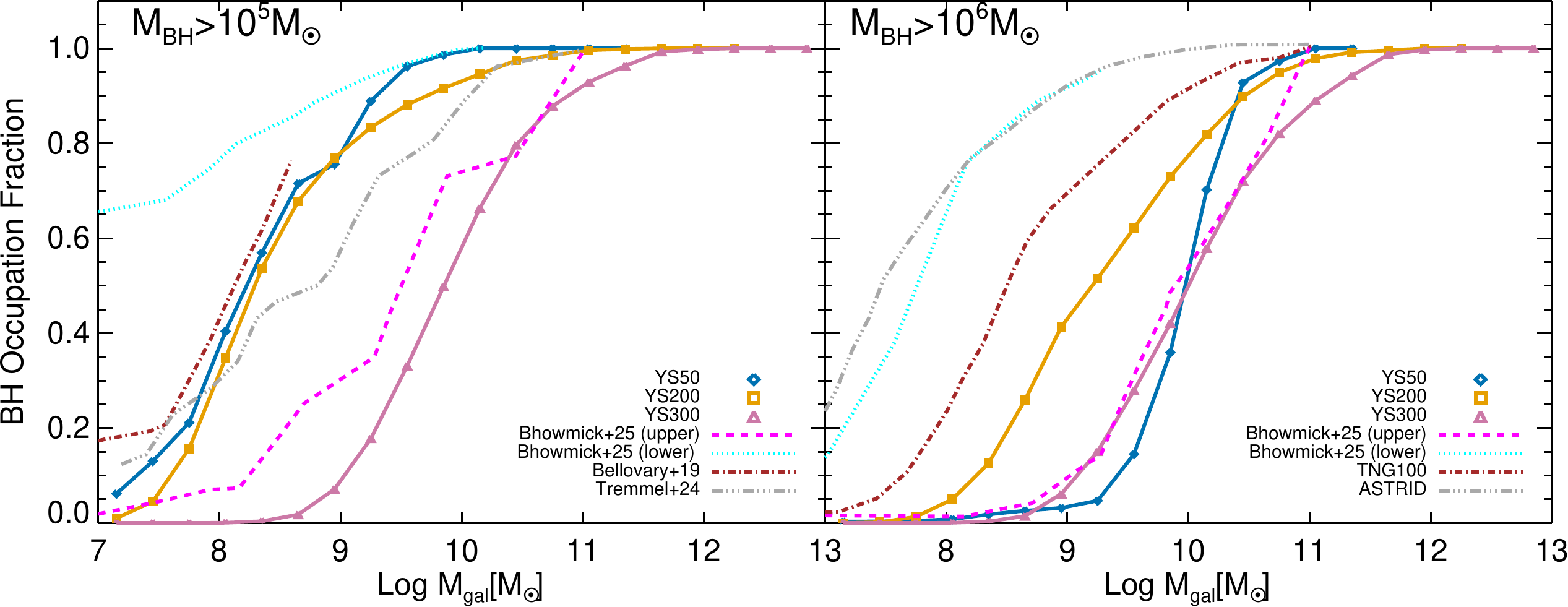}
\caption{Comparison between the black hole occupation fraction, $f_{\rm BH,occ}$, predicted by our model in the three simulation boxes and results from observational studies (upper panels) and numerical simulations (lower panels). The left and right columns correspond to $M_{\rm BH}>10^5\,M_{\odot}$ and $M_{\rm BH}>10^6\,M_{\odot}$, respectively. Our predictions for YS50, YS200, and YS300 are shown in blue diamonds, orange squares, and purple triangles, respectively. For the lower BH mass threshold, YS50 and YS200 lie close to the range inferred by \citet{miller2015} and show a broader overlap with the constraints of \citet{zou2025}, while both predict lower occupation fractions than those inferred by \citet{burke2025}. For the higher threshold, YS200 lies closest to the observational estimates of \citet{greene2012}, whereas YS50 and YS300 predict lower occupation fractions over most of the stellar-mass range probed
by the data. In comparison with numerical predictions, YS50 and YS200 broadly overlap with the range spanned by \citet{bhowmick2025}, \citet{bellovary2019}, and \citet{tremmel2024} for the lower threshold. For the higher threshold, YS200 lies within the range covered by several models and is closer to the TNG100 prediction \citep{nelson2019} than to ASTRID \citep{ni2025}, while YS300 shows the largest overall offset from the literature results. The observational comparison should be regarded as illustrative, since the literature constraints are not necessarily based on
the same sharp BH mass thresholds adopted in our model and may depend on
luminosity limits, duty cycle, and selection effects.}
\label{fig:fig7}
\end{figure*}

To conclude our analysis, we compare our $z=0$ predictions with a wide set of observational and numerical studies. This comparison is shown in Figure~\ref{fig:fig7}, where we plot the $f_{\rm BH,occ}-M_{\rm gal}$ relation predicted by our model for YS50 (blue diamonds), YS200 (orange squares), and YS300 (purple triangles). The upper panels show the comparison with observational constraints for the two adopted BH mass thresholds, while the lower panels show the comparison with results from numerical simulations.

Before discussing the comparison, we briefly summarize the nature of the literature data shown in Figure~\ref{fig:fig7}. The observational constraints are not all based on the same selection: some are inferred
from accretion signatures and therefore depend on luminosity limits, duty cycle, and accretion-rate distributions, while others attempt to constrain the intrinsic occupation fraction more directly. The numerical
predictions also differ in seeding prescriptions, feedback models, resolution, and galaxy-selection criteria. The observational data should therefore be interpreted as approximate comparisons, rather than as measurements selected through the same sharp BH mass thresholds adopted in our model.

We first focus on the observational comparison. For the lower BH mass threshold, $M_{\rm BH}>10^5 \,{\rm M}_{\odot}$, YS50 and YS200 lie close to the range inferred by \citet{miller2015}, and show a somewhat weaker
but still broadly consistent overlap with \citet{zou2025}. Our predictions instead lie below the high occupation fractions inferred by \citet{burke2025}. This difference is particularly informative because the constraints of \citet{burke2025} rely on a multiwavelength inference of the intrinsic occupation fraction and depend on assumptions about the accretion-rate distribution and active fraction. This highlights that
observationally inferred occupation fractions may correspond to an effective, model-dependent mapping between luminosity-selected AGN samples and the underlying BH population.

When the higher BH mass threshold adopted in this study is considered, $M_{\rm BH}>10^6 \,{\rm M}_{\odot}$, the predictions from YS200 are the closest to the observational estimates of \citet{greene2012}. YS50 and
YS300 instead predict lower occupation fractions over most of the stellar-mass range probed by the data. However, this comparison should also be interpreted with caution, because the observational estimates are not necessarily selected through a BH mass threshold identical to the one adopted here.

We stress that the comparison with observational constraints should be interpreted with caution. While our model measures the intrinsic occupation fraction above a fixed BH mass threshold, observational estimates are generally based on accretion signatures and are therefore sensitive to luminosity limits, Eddington-ratio distributions, duty cycle, and contamination from stellar processes. Discrepancies with individual observational curves do not necessarily imply a failure of the model, but may partly reflect differences in the operational definition of BH occupation.

In the lower panels, we compare our predictions with results from numerical simulations. For the lower BH mass threshold, YS50 and YS200 are very close to each other and lie close to the predictions of \citet{bhowmick2025}, \citet{bellovary2019}, and \citet{tremmel2024}. YS300, instead, lies systematically below most of the simulation results. Similar trends are found when the BH mass threshold is increased. In
this case, the overlap between YS50 and the other simulations becomes weaker, while YS200 lies within the range spanned by the models of \citet{bhowmick2025}, remains offset from ASTRID \citep{ni2025}, and is
somewhat closer to TNG100 \citep{nelson2019}. Once again, among our three boxes, YS300 shows the largest deviation from the other predictions.

The comparison with theoretical studies reinforces the interpretation developed in the previous sections. At the lower BH mass threshold, YS50 and YS200 broadly overlap with the range spanned by other simulations, suggesting that our model predicts a plausible occupation fraction in the low-mass regime when the host population is sufficiently resolved. At the higher threshold, the agreement becomes more model-dependent, with YS200 lying closer to the range covered by \citet{bhowmick2025}, TNG100, and ASTRID, while YS300 remains systematically shifted toward higher stellar masses. This supports the idea that the inferred occupation fraction is highly sensitive to the combination of BH mass threshold, numerical resolution, and sampled galaxy population. The recent BRAHMA results of \citet{bhowmick2026} further illustrate the wide range of local occupation fractions produced by different seed models and BH dynamics prescriptions.

Overall, these comparisons show that the $f_{\rm BH,occ}-M_{\rm gal}$ relation, both observationally and theoretically, is still not fully understood. This is due not only to the intrinsic difficulty of measuring BH masses observationally and of modeling BH seeds numerically, but also to the role played by environment, galaxy selection, and, in the case of simulations and SAMs, numerical resolution. With these caveats in mind, we now conclude by summarizing the main results of this study.

\section{Conclusions}
\label{sec:conclusion}

We take advantage of our state-of-the-art SAM \texttt{FEGA25}, run on the merger trees of three dark matter-only cosmological simulations, to study the relation between the BH occupation fraction ($f_{\rm BH,occ}$) and galaxy stellar mass. We investigate this relation as a function of BH mass threshold, galaxy type, simulated volume, numerical resolution, sampled galaxy population, and redshift.

Our analysis focuses exclusively on this relation and explores several key aspects that can provide insight into the growth of BHs over cosmic time. In particular, we highlight which factors should be carefully considered when comparing different results, both theoretical and observational, and, in the case of numerical methods, the role played by the simulated volume in determining the inferred BH occupation fraction.

In the following, we briefly summarize our main conclusions:
\begin{itemize}
 \item Our model \texttt{FEGA25} is able to reproduce the observed BH mass function from at least $z=2$ to $z=0$. This is an important consistency check given the main goal of this study.

 \item There is a clear distinction between the $f_{\rm BH,occ}$-stellar mass relations of central and satellite galaxies, although their relative behavior depends on the simulation box
 and on the adopted BH mass threshold. In YS50, satellites have slightly higher occupation fractions than centrals for the lower threshold, while the two populations show similar relations for the higher threshold. In YS200 and YS300, central galaxies are already massive enough to produce $f_{\rm BH,occ}\sim1$ over most of the sampled stellar-mass range, whereas the satellite occupation fraction increases more gradually with stellar mass.

 \item The global $f_{\rm BH,occ}$-stellar mass relation increases monotonically with stellar mass, but its normalization and shape should be interpreted as population-weighted quantities. They depend on the  occupation fractions of central and satellite galaxies separately, as well as on the relative abundance of the two populations in each stellar-mass bin. Because the occupation fractions of central and satellite galaxies differ in a box- and threshold-dependent way, changes in their relative abundance, evolutionary histories, and retained BH masses can significantly affect the inferred global relation.

 \item The predicted occupation fraction is sensitive to box-to-box variations, reflecting the combined impact of numerical resolution, simulated volume, and sampled galaxy population. The highest-resolution  simulation, YS50, resolves lower-mass halos and their associated BHs more effectively, while the lower-resolution boxes tend to shift the transition to high occupation fractions toward larger stellar masses.  At the same time, larger volumes sample a wider range of environments, including groups and clusters, thereby changing the relative contribution of central and satellite galaxies.

 \item The redshift evolution of the $f_{\rm BH,occ}$--stellar mass relation is not universal. YS50 and the \texttt{NewCluster} zoom-in simulation show higher occupation fractions at higher redshift, in qualitative agreement with the trend found in Romulus25/RomulusC by \citet{tremmel2024}. By contrast, the larger-volume boxes show the opposite behavior. This suggests that the inferred redshift evolution is sensitive to the galaxy population sampled at each epoch, and may reflect a combination of environment, numerical resolution, satellite fraction, and progenitor bias, rather than a single evolutionary track at fixed stellar mass.

 \item The relation predicted by our model broadly overlaps with the range spanned by several observational estimates and numerical predictions, including hydrodynamical simulations with different resolutions and  volumes, as well as other SAMs. However, the inferred occupation fraction is highly sensitive to the combination of BH mass threshold, simulated volume, numerical resolution, and sampled galaxy population. Comparisons with observational constraints must therefore be interpreted with caution, because our model measures an intrinsic occupation fraction above a fixed BH mass threshold, whereas observations generally infer BH occupation through accretion signatures and are sensitive to luminosity limits, duty cycle, Eddington-ratio distributions, and contamination from stellar processes.
\end{itemize}

Taken together, our results show that $f_{\rm BH,occ}$ is not a universal function of galaxy stellar mass alone. Instead, its inferred shape and normalization depend on the adopted BH mass threshold, the central or satellite nature of the host galaxy, the numerical resolution of the simulation, and the galaxy population sampled by a given volume. This has important implications for both theoretical and observational studies. On the theoretical side, robust predictions require sufficient resolution to follow low-mass galaxies and their BHs, as well as a careful treatment of the relative contribution of satellites. On the observational side, meaningful comparisons require a clear connection between intrinsic occupation fractions and luminosity-selected samples of actively accreting BHs. Future progress will therefore rely on combining higher-resolution simulations, larger and more complete galaxy samples, and forward-modeled observational selections, in order to place stronger constraints on the origin and evolution of BHs in low-mass galaxies.


\section*{Acknowledgements}
The authors thank the referee for his very constructive comments which helped to improve the manuscript.
E.C. and S.K.Y. acknowledge support from the Korean National Research Foundation (RS-2025-00514475). E.C. acknowledges support from the Korean National Research Foundation (RS-2023-00241934). E.C., J.K.J., C.S., and S.K.Y. are supported by the Korean National Research Foundation (RS-2022-NR070872). J.R. acknowledges support by the Institut de Physique des deux infinis of Sorbonne Université and by the ANR grant ANR-19-CE31-0017 of the French Agence Nationale de la Recherche.

\section*{Data Availability}
The derived data products used to produce the figures and results presented in this work are available from the corresponding author upon reasonable request. This includes the processed quantities underlying the figures and the galaxy catalog information required to reproduce the analyses presented here. Requests should be directed to the corresponding author and
should specify the data products required. The full simulation outputs (\texttt{NewCluster}) are publicly available at \url{https://gemsimulation.com}.

\appendix

\setcounter{figure}{0}
\renewcommand{\thefigure}{A\arabic{figure}}

\section{Black Hole Mass Function Across the Simulation Boxes}
\label{app:bhmf}

As discussed in the main text, the three dark matter simulations used in this work differ substantially in simulated volume, mass resolution, and in the galaxy populations that they sample. Since these differences may also affect the predicted global BH population, we compare here the BH mass functions obtained with \texttt{FEGA25} when applied independently to YS50, YS200,
and YS300. The mass functions are computed consistently using the full available volume of each simulation and adopting the same stellar-mass selection, BH-mass binning, and normalization procedure in all three cases. This provides a complementary view of the simulation-to-simulation variations discussed in the main analysis and allows us to assess their impact on the global BH population independently of the occupation fraction.

\begin{figure*}
    \centering
    \includegraphics[width=0.90\textwidth]{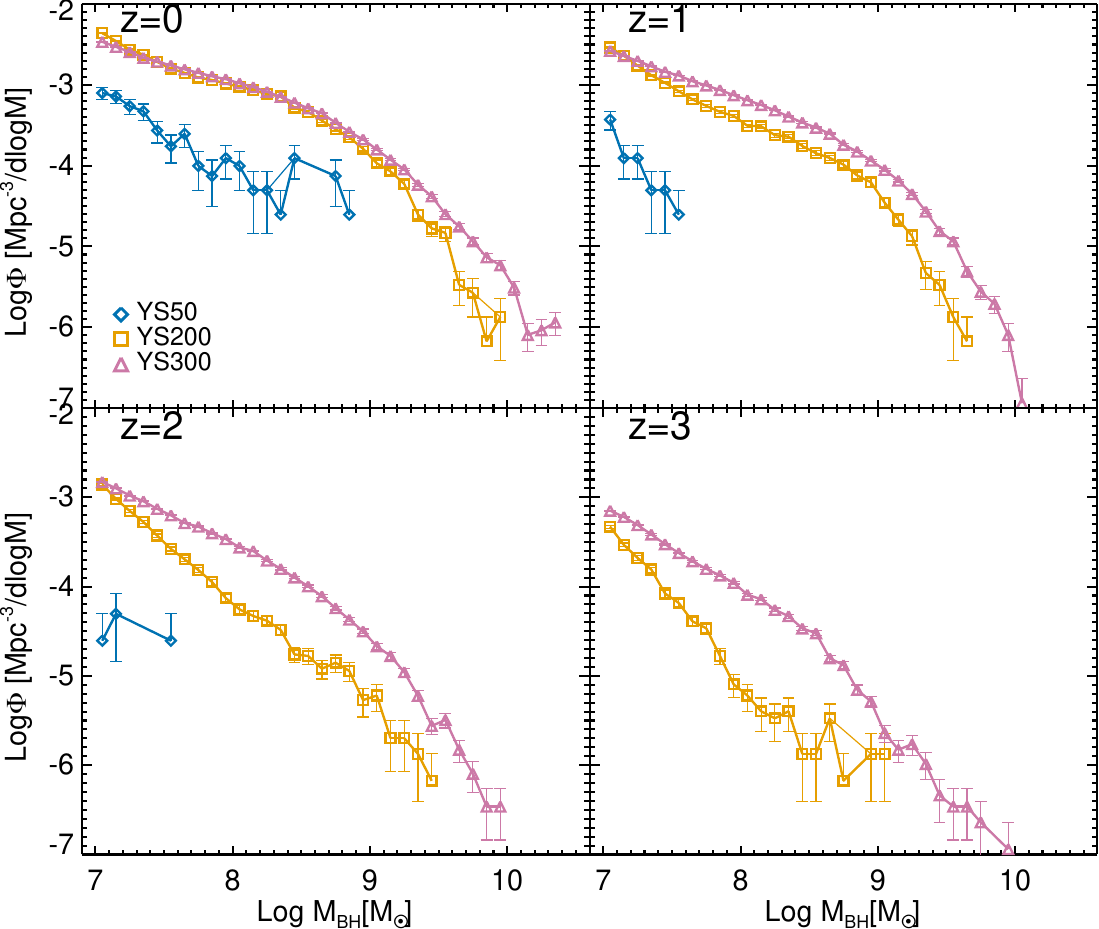}
    \caption{Black hole mass functions predicted by \texttt{FEGA25} using YS50, YS200, and YS300 at $z=0$, $1$, $2$, and $3$, as indicated in the individual panels. Different colors and symbols identify the three simulation boxes: YS50 (blue diamonds), YS200 (orange squares), and YS300 (purple triangles). The mass functions are computed over the full available volume of each simulation. The comparison illustrates the
    simulation-to-simulation variation of the predicted BH population associated with the different mass resolutions, simulated volumes, and galaxy populations sampled by the three boxes.}
    \label{fig:bhmf_boxes}
\end{figure*}

Figure~\ref{fig:bhmf_boxes} shows appreciable box-to-box variations in the predicted BH mass function. YS200 and YS300 give broadly similar number densities over much of the BH-mass range sampled by both simulations, although some differences remain and become more evident toward higher redshift. YS50 instead predicts systematically lower abundances and samples a more limited range of BH masses, with the difference becoming progressively more pronounced at increasing redshift.

Since the mass functions are computed consistently over the full available volume of each simulation, these differences cannot be attributed to the volume normalization alone. Changing the simulation box simultaneously changes the mass resolution, the statistical sampling of rare systems, and the halo populations and assembly histories represented by the corresponding merger trees. In particular, the smaller volume of YS50 provides a more limited sampling of rare massive systems, while its higher mass resolution allows lower-mass structures to be followed in greater detail. Conversely, YS300 samples a substantially larger cosmological volume and therefore a larger population of massive halos and dense environments, but at lower mass resolution. YS200 represents an intermediate case between these two regimes.

The increasing differences toward higher redshift are therefore likely to reflect the greater sensitivity of the rarer BH population to the combined effects of volume, mass resolution, and population sampling. At these epochs, massive BHs and their progenitors are intrinsically less common, making their predicted abundance more sensitive to the particular halo populations and assembly histories represented in each simulation. We therefore interpret Figure~\ref{fig:bhmf_boxes} as illustrating the box-to-box variation of the predicted global BH population rather than as a numerical convergence test. This comparison provides additional context for the corresponding simulation-to-simulation variations in the BH occupation fraction discussed in the main text.

\bibliography{paper}{}
\bibliographystyle{aasjournalv7}



\end{document}